\documentclass[%
      aps,
      prb, 
      nobibnotes,
      twocolumn,
      longbibliography,
      superscriptaddress,
      showkeys]{revtex4-1}
      
\usepackage[utf8]{inputenc}
\usepackage[T1]{fontenc}
\usepackage{dcolumn}
\usepackage{bm}					
\usepackage{graphicx}
\usepackage{amsmath, amsfonts, amssymb, mathtools}	
\usepackage{latexsym} 				
\usepackage{xcolor}
\usepackage[caption = false]{subfig}
\usepackage[breaklinks, linkcolor = black]{hyperref}
\usepackage{tikz}
\usepackage{tabularx}
\usepackage{braket}
\usepackage{diagbox}
\usepackage[normalem]{ulem}
\usepackage{textcomp}
\usepackage{soul}
\usepackage{booktabs,eqparbox}
\usepackage{comment} 

\begin{document}

\title{Spin-orbit torque in single-molecule junctions from \textit{ab initio}} 

\author{Mar\'ia Camarasa-G\'{o}mez}

\email[e-mail:]{maria.camarasa-gomez@weizmann.ac.il}

\affiliation{Institute of Theoretical Physics, University of Regensburg, 93040 Regensburg, Germany}

\affiliation{Department of Molecular Chemistry and Materials Science, Weizmann Institute of Science, Rehovot 7610001, Israel}

\author{Daniel \surname{Hernang\'{o}mez-P\'{e}rez}}

\affiliation{Institute of Theoretical Physics, University of Regensburg, 93040 Regensburg, Germany}

\affiliation{CIC nanoGUNE BRTA, Tolosa Hiribidea 76, 20018 San Sebasti\'an, Spain}

\author{Ferdinand Evers}

\email[e-mail:]{ferdinand.evers@ur.de}

\affiliation{Institute of Theoretical Physics, University of Regensburg, 93040 Regensburg, Germany}

\date{\today}

\begin{abstract}
\subparagraph{ABSTRACT:} The use of electric fields applied across magnetic heterojunctions that 
lack spatial inversion symmetry has been previously proposed as a non-magnetic mean of controlling localized magnetic moments through spin-orbit torques (SOT). The implementation of this concept at the single-molecule level has remained a challenge, however. 
Here, we present first-principle calculations of SOT in a single-molecule junction under bias and beyond linear response. Employing a self-consistency scheme invoking density functional theory and non-equilibrium Green's function theory, we compute the current-induced SOT. Responding to this torque, a localized  magnetic moment can tilt. 
Within the linear 
regime our quantitative estimates for the SOT in single-molecule junctions yield values similar to those known for magnetic interfaces. Our findings contribute to an improved microscopic understanding of SOT in single molecules. 
\end{abstract} 


\maketitle

\subparagraph{Introduction.} The controlled manipulation of magnetic moments in solid state heterojunctions by driving a spin-polarized electrical current is an established technique in spintronics\cite{Zutic2004} since its theoretical prediction almost three decades ago\cite{Slonczewski1996, Berger1996} and its subsequent experimental observation\cite{Tsoi1998, Sun1999}. This type of spin-transfer torque (STT) \cite{Stiles2002, Stiles2008} can be used to produce current-induced magnetization reversal even at room temperature\cite{Buhrman1999, Katine2000, Albert2002, Ralph2003, Stiles2008, Liu2012} by transfer of spin angular momentum between the charge carriers and the localized magnetic moments; 
applications include STT random-access memories \cite{Bhatti2017}.
A potential drawback of STT-based technology is the involvement of high density spin-polarized currents\cite{Saha2022}. Such currents require balancing of magnetoresistance, cross section and bias voltage, which is believed to be challenging for downsizing devices\cite{Gambardella2011, Song2021}.

A promising effect that can produce current-induced magnetization switches all electrically, i.e. without invoking spin-polarized driving currents, \cite{Prenat2015, Seo2016, Bhatti2017, Sato2018, Honjo2019, Be2019, Shao2021, Song2021, Ryu2022} is the spin-orbit torque\cite{Chernyshov2009, Miron2010, Gambardella2011, Gambardella2019, Song2021} (SOT); here, the exchange of angular momentum between a local magnetization and the driving current is mediated by spin-orbit (SO) interaction. SOT on localized moments can be produced in systems {lacking} structure inversion symmetry \cite{Manchon2009, Garate2009}. 
Furthermore, due to the potentially small current densities required, it opens up the opportunity to scale down devices to the molecular or even atomic level.

In this work we devise a self-consistent theoretical scheme to study the SOT in a molecular junction. 
The simulation method we propose extends previous studies that have confined themselves to the linear regime, i.e. a Kubo-formula  based calculation of the torkance tensor\cite{Freimuth2014, Geranton2015}. Similar to the earlier studies, we also adopt the adiabatic approximation, in which the Hamiltonian is assumed to be time-independent\cite{Haney2007}. 
Our non-equilibrium Green's function (NEGF) approach is based on density functional theory (DFT); it therefore incorporates details of the electronic structure and geometry, thus extending previous  model-based investigations\cite{Manchon2009, Kwaku2014}. 
\begin{figure}[b]
    \includegraphics[width=0.70\linewidth]{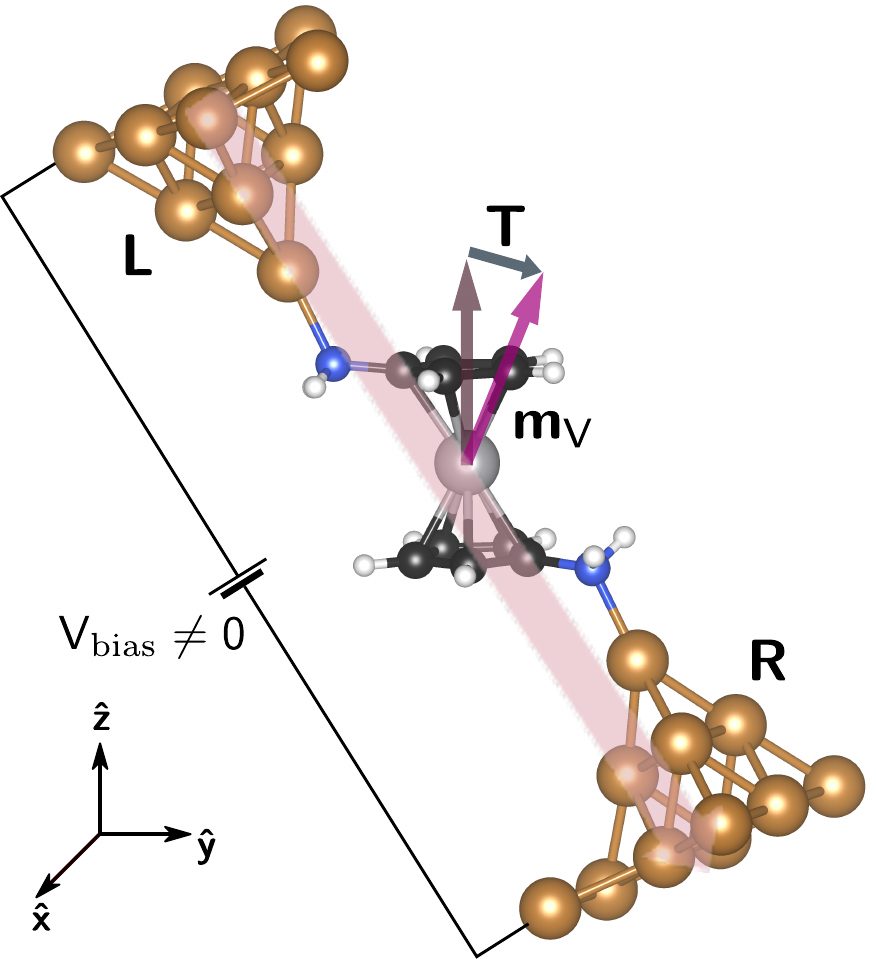}      
     \caption{Schematic structure of the molecular junction. As a response to a voltage bias, $V_\textnormal{bias}$, the current, represented by a semi-transparent red arrow, flows from the left (L) to the right (R) contact. The magnetic moment, located mostly at the vanadium atom and labeled by $\mathbf{m}_V$, reacts and tilts by means of a spin-orbit torque, $\mathbf{T}$, associated with the current flow.
      }\label{f1} 
\end{figure}

As a case study, we apply our simulation method to a vanadocene-copper molecular junction, Fig. \ref{f1}.  
Vanadocene is an open-shell metallocene \cite{Long1998}, formed by a vanadium atom ``sandwiched'' between two carbon-based cyclopentadienyl (Cp) rings in a double decker structure. Some of the compounds of this family have recently sparked a renewed interest in view of their intrinsic quantum interference\cite{Camarasa2020} and light-induced properties\cite{Lee2024}.
We find that an all electrical  control of the localized magnetic moment can be achieved. Within the linear regime, we extract from our results values for the torkance that are quantitatively similar to the previous estimates based on the Kubo-formula.  

 \begin{figure}
     \includegraphics[width=1.0\linewidth]{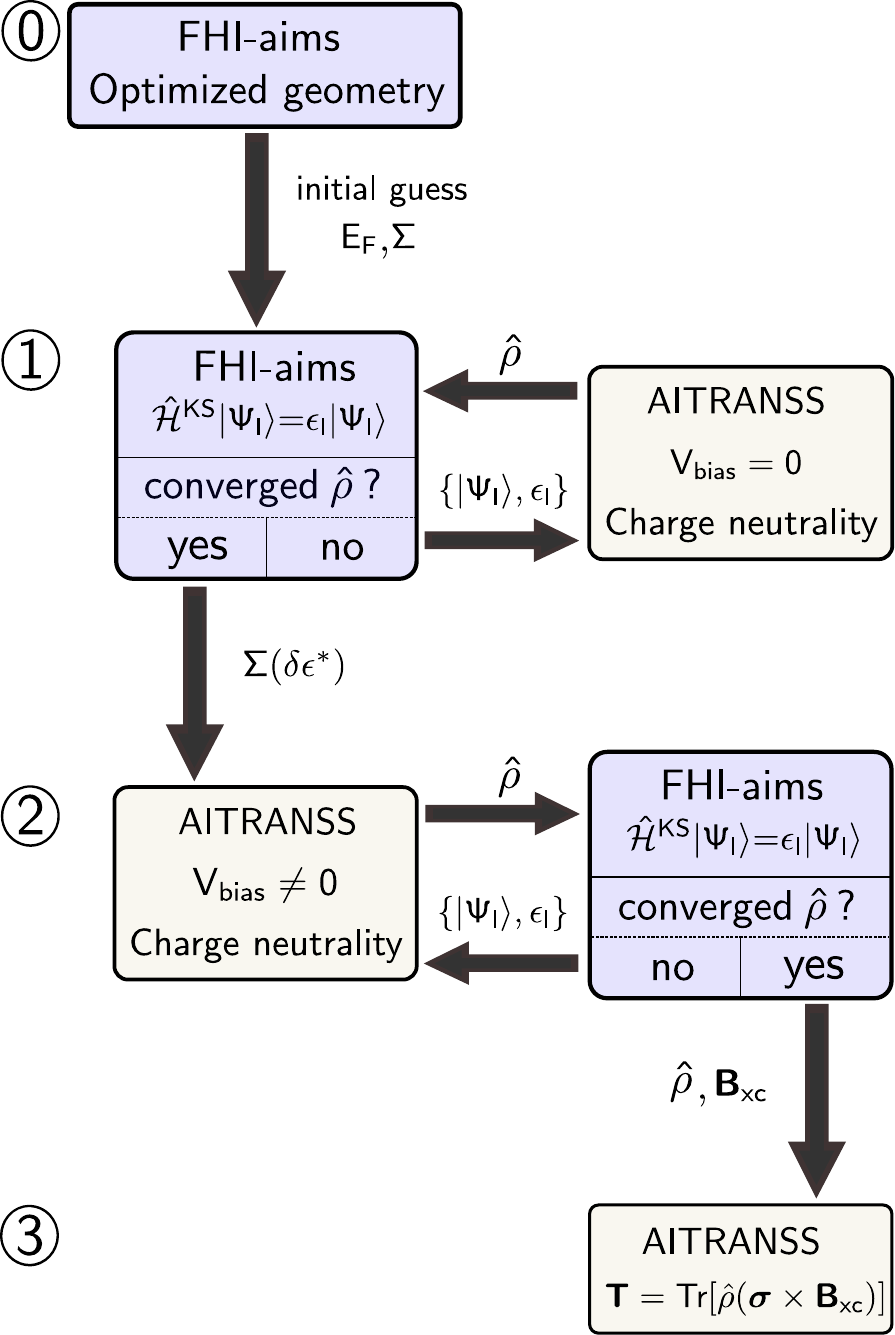}      
    \caption{Schematic of the self-consistent cycle considered in this work. The calculation proceeds in four sequential steps (labeled from \textcircled{\raisebox{-0.9pt}{0}}-\textcircled{\raisebox{-0.9pt}{3}}). After an initial geometry optimization in step \textcircled{\raisebox{-0.9pt}{0}}, the self-energy, $\Sigma$, and the Fermi energy, $E_\textnormal{F}$ are parametrized in step \textcircled{\raisebox{-0.9pt}{1}}. The optimal self energy for a given Fermi energy and real self-energy shifts, $\Sigma(\delta \epsilon^\ast)$, is used in the second self-consistent loop at finite bias voltage, see step \textcircled{\raisebox{-0.9pt}{2}}. A further postprocessing step allows to compute physical observables with the optimized non-equilibrium density matrix, see step \textcircled{\raisebox{-0.9pt}{3}}.
    Additional details of the cycle are described in the SI.
     }\label{f2} 
 \end{figure}

{\bf Model.} Our simulations operate within the framework of non-collinear DFT with a corresponding Kohn-Sham Hamiltonian\cite{Capelle2001, MacDonald2008, Nikolic2018, Vignale1989, Ullrich2018}
\begin{equation}\label{eq:Ham_nc}
    \hat{\mathcal{H}}^{\textnormal{KS}}(\mathbf{r}) = \hat{\mathcal{H}}^{(0)}(\mathbf{r}) \otimes 1\!\!1_2  + \bm{\sigma} \cdot
     \hat{\mathbf{B}}_{\textnormal{xc}}(\mathbf{r}),
\end{equation}
with $\bm{\sigma}$ the vector of Pauli matrices, $\bm{\sigma} = (\sigma_{x}, \sigma_{y}, \sigma_{z})$; $1\!\!1_2$  the $2 \times 2$ identity matrix, $\hat{\mathcal{H}}^{0}$  the spin-independent part of the Kohn-Sham Hamiltonian and $\hat{\mathbf{B}}_{\textnormal{xc}}$ the exchange-correlation (XC) magnetic field embodying the effect of SO coupling. 
The XC magnetic field is formally given by the functional derivative of the XC energy with respect to the magnetization, $\hat{\mathbf{B}}_\textnormal{xc}(\mathbf{r}) = \delta E_{\textnormal{xc}}[n(\mathbf{r}), \mathbf{m}(\mathbf{r})]/\delta \mathbf{m}(\mathbf{r})$. 

{\bf Nonequilibrium formalism.} Using the Kohn-Sham states, $|\Psi_l \rangle$, and energies, $\epsilon_l$, 
of Eq. \eqref{eq:Ham_nc} 
we build the non-equilibrium density matrix, $\hat{\rho} (V_\textnormal{bias})$ following the general construction rules as formulated in Refs. \onlinecite{Arnold2007,Evers2011,Evers2020}.  
The finite-bias expectation values of observables, $\hat O$,  are calculated as usual, i.e. by evaluating the trace  $O \coloneqq\langle \hat O \rangle = \textnormal{Tr}[\hat{\rho}\,\hat{O}]$. This way we have for the total particle number, $N_\textnormal{elec} \coloneqq \langle N \rangle =  \textnormal{Tr}(\hat \rho \hat N)$, the (spin) magnetization, $\mathbf{m} \coloneqq \langle \hat{\mathbf{m}}\rangle = \mu_B  \textnormal{Tr} (\hat \rho \bm{\sigma})$ [with Bohr's magneton $\mu_B = e\hbar/(2m)$] and for the SOT 
\begin{align}
\mathbf{T} \coloneqq  \langle \mathbf{\hat{T}}  \rangle &= \textnormal{Tr}\left[\hat{\rho} (\bm{\sigma}\times \hat{\mathbf{B}}_{\textnormal{xc}})\right]
\label{e2}\\
&=\textnormal{Tr}\left[\delta\hat{\rho} (\bm{\sigma}\times \hat{\mathbf{B}}_{\textnormal{xc}})\right]
\label{e3}
\end{align} 
where $\delta\hat{\rho}\coloneqq\hat{\rho}(V_\text{bias})-\hat{\rho}(0)$. The second line reflects the fact that in equilibrium
 by definition all physical observables are stationary and therefore the SOT has to vanish, $ \mathbf{T}^{\textnormal{eq}} = \textnormal{Tr}[\hat{\rho}(0)(\bm{\sigma} \times \hat{\mathbf{B}}_{\textnormal{xc}})] = \mathbf{0}$ ( ``zero-torque theorem'' \cite{Capelle2001, Scuseria2013}).
\footnote{We note that equilibrium requires only the trace \eqref{e2} to vanish; the individual matrix elements that constitute the trace do not vanish by themselves. The observation can be relevant if the commutator $[\hat{\rho}(0),\hat{\mathcal{H}}]$ 
does not vanish. This can happen as an an artifact of the finite simulation volume; in this case the two expressions \eqref{e2} and \eqref{e3} are not equivalent and \eqref{e3} is to be preferred. 
The artifact is controlled by the size of the simulation volume and vanishes in the thermodynamic limit.} 
We further mention that the form of the operator 
$\mathbf{\hat{T}}$ reflects the rate of change of the spin density due to the local exchange-correlation field created by the SO interaction, $\hat{\mathbf{T}} = {d\hat{\mathbf{S}}}/{dt}$ where  $\hat{\mathbf{S}}={(\hbar/2)}\bm{\sigma}$.\cite{Nikolic2018} Indeed, using Heisenberg's equation of motion, one has  
\begin{equation}\label{torqueopdev}
    \hat{\mathbf{T}} =\frac{1}{2i}[\bm{\sigma},\hat{\mathcal{H}}^{\textnormal{KS}}] =   \bm{\sigma}\times\hat{\mathbf{B}}_{\textnormal{xc}}.
\end{equation}

{\bf Self-consistent simulation procedure.} We display in Fig. \ref{f2} the non-equilibrium self-consistency cycle; the  technical details have been relegated to the Supporting Information (SI). In short, starting from an optimized geometry of the molecular junction obtained with the FHI-aims package\cite{Blum2009}, we perform a self-consistent DFT\cite{Dreizler1990, Parr1989}--NEGF\cite{DiVentra2008, Cuevas2010} cycle  that accounts for the charge redistribution in the junction due to the macroscopic nature of the contacts\cite{Arnold2007, Bagrets2013, Camarasa2021, Garcia-Blazquez2023, Naskar2023}. We ensure the charge neutrality and screening the excess charge accumulated at the boundaries of the finite cluster by a self-energy model implemented in the module AITRANSS\cite{Arnold2007, Evers2011, Bagrets2013}. This module was recently extended to include SO coupling\cite{Camarasa2021}. 

Typically, the spin carried by the molecular junction implies only a weak magnetic anisotropy energy (MAE) and, as a consequence, the self-consistent field (SCF) iterations are difficult to converge. 
As is often done in situations with a weakly broken continuous symmetry, we stabilize the convergence of the scf-cycle by  operating with a manually enhanced  MAE  
quantified by a small parameter $\Delta$ (see SI for further details); at the end of the calculations we extrapolate into the limit $\Delta\to 0$. Since $\Delta$ is small, the (weakly) enhanced MAE does not alter the orbital ordering; the corresponding shift of the Kohn-Sham energies is less than the average level spacing close to the frontier orbitals and therefore negligible. 

As a demonstration of our implementation of the SCF cycle for the SOT, we consider the junction Fig. 
\ref{f1}, i.e. a single vanadocene molecule\cite{Prins2004} between two copper contacts. Copper electrodes recommend themselves due to their weak SO interaction.

{\bf Reference calculation - isolated molecule} Vanadocene possesses a relatively small SO interaction, with uniaxial anisotropy and an easy magnetization axis perpendicular to the plane containing each of the Cp rings. Most of the magnetic moment is located on the vanadium atom and is associated with three unpaired electrons in a $e_{2g} + a_{1g}$ configuration\cite{Prins2003}. Consequently, DFT (with the
PBE functional\cite{Perdew1996})
 yields 3.0 $\mu_\text{B}$ for the spin magnetic moment  in the isolated molecule, with the spin of the molecule oriented in the $\hat{\mathbf{z}}$ direction. 
Further, the DFT-calculation confirms a weak MAE, smaller than 1 meV, explaining the need of using the MAE enhanced by $\Delta$ for stabilization of the SCF cycle when bringing the molecule in contact with the electrodes.

\begin{figure}[t]
   \includegraphics[width=1.0\linewidth]{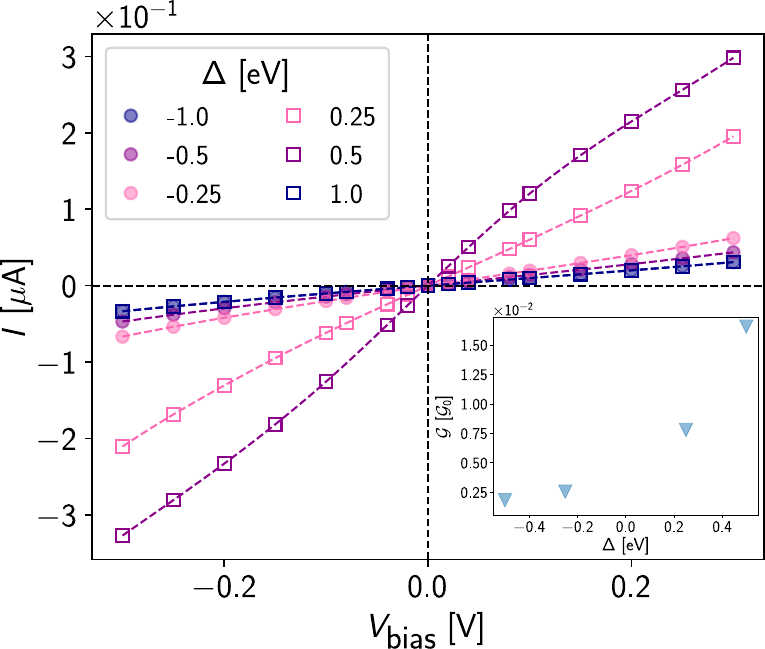}
    \caption{Current-voltage ($I-V_\textnormal{bias}$) characteristics of the vanadocene junction for several values of the magnetic anisotropy parameter, $\Delta$. Dashed lines cubic-spline interpolations.
    Inset: (Zero-bias)
    conductance $\mathcal{G}$ (in units of the conductance quantum) extracted from the current-voltage characteristics as a function of $\Delta$ (dashed lines are guide to the eye) from which the limit $\mathcal{G}(\Delta \rightarrow 0)$ can be inferred.
    }\label{f3} 
\end{figure}

{\bf Results - transport.} In Fig. \ref{f3}, we show the current-voltage characteristics  ($I-V_\textnormal{bias}$) of the vanadocene-copper junction obtained after employing the self-consistent cycle in Fig. \ref{f2}. We observe that the $I-V_\textnormal{bias}$ is linear up to $V_\textnormal{bias} \sim 40 $ meV, above which deviations due to higher-order terms in the applied bias are found; we associate them to polarization effects that have been discussed in Ref. \onlinecite{Evers2020}. These effects are largely independent of the MAE in the junction. We also display in the inset the zero-bias conductance, $\mathcal{G}$, which can be extracted from the $\textnormal{d}I/\textnormal{d}V_\textnormal{bias}$ in the limit $V_\textnormal{bias} \rightarrow 0$. Taking the limit $\mathcal{G}(\Delta \rightarrow 0)$ we find $\sim 5 \cdot 10^{-3}$ (in units of the conductance quantum, $\mathcal{G}_0 = 2e^2/h$). This result is consistent with  transmission calculations performed for ferrocene (amine-linked) gold molecular junctions\cite{Camarasa2020}.

\begin{figure}[t]
    \includegraphics[width=1.0\linewidth]{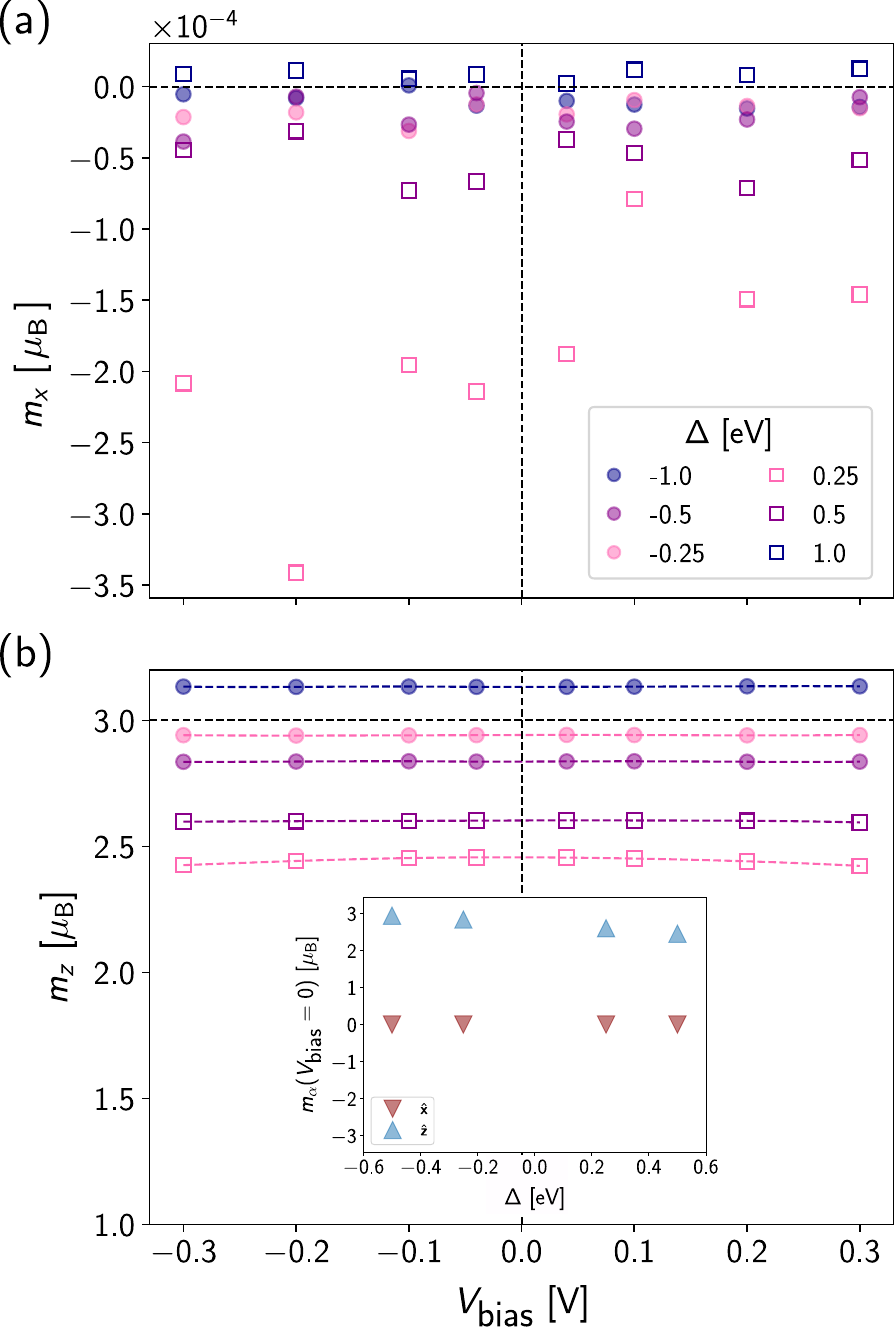}  
    \caption{Magnetization, $m_\alpha$, at the vanadium atom as a function of the voltage bias, $V_\textnormal{bias}$, applied across the molecular junction. We display the traces obtained for several representative values of the magnetic anisotropy term in the two relevant spatial directions (a) $\hat{\mathbf{x}}$ and (b) $\hat{\mathbf{z}}$. The dashed horizontal line represents the value expected for the local magnetization in the isolated vanadocene molecule, oriented in the $\hat{\mathbf{z}}$ direction. The  dashed colored lines correspond to cubic-spline interpolations. Inset: Zero-bias magnetization  as a function of $\Delta$.
  }\label{f4} 
\end{figure}
{\bf Results - magnetism.} 
In Fig. \ref{f4}, we display the local magnetic moment, $\mathbf{m}$, at the position of the vanadium atom as a function of $V_\textnormal{bias}$. We show the magnetic moment in (a) the $\hat{\mathbf{x}}$ and (b) the $\hat{\mathbf{z}}$ direction (see SI for the $\hat{\mathbf{y}}$ direction) considering as well several values for the magnetic stabilization term, $\Delta$. 
As  a reference, we also display as a dashed horizontal line the value obtained for the local magnetization in the isolated vanadocene molecule, which is directed towards the easy axis ($\hat{\mathbf{z}}$ direction).
We observe first that also after the electrodes have been attached, the magnetic moment is mainly carried by the vanadocene atom, with values close to $\sim 3.0\mu_\text{B}$ and points in the $\hat{\mathbf{z}}$ direction.  Also, in both $\hat{\mathbf{x}}$ and $\hat{\mathbf{z}}$ directions, the zero-bias value is largely insensitive to the value of $\Delta$ (see inset).
We note that $m_x$ exhibits large fluctuations (``noise'') with the applied voltage. Given the small absolute values of order $10^{-4}\mu_\text{B}$, we 
attribute  the noise to limits of the SCF convergence.

Furthermore, we find that $m_z$ presents a small curvature corresponding to smooth variations of the order of $\sim 1$\% when the current flows across the molecular junction. In this sense, within the accuracy of the calculation $m_z$ is not strongly dependent on the current flow because the spin is strongly locked into the easy axis and the SO interaction, which drives the SO torque, is small.
We interpret the curvature of $m_z$ with the applied V\textsubscript{bias} as the result of transfer of angular momentum of the itinerant electrons associated to the stationary current flow to the localized spins due to the SO torque.
%

\begin{figure}
    \includegraphics[width=1.0\linewidth]{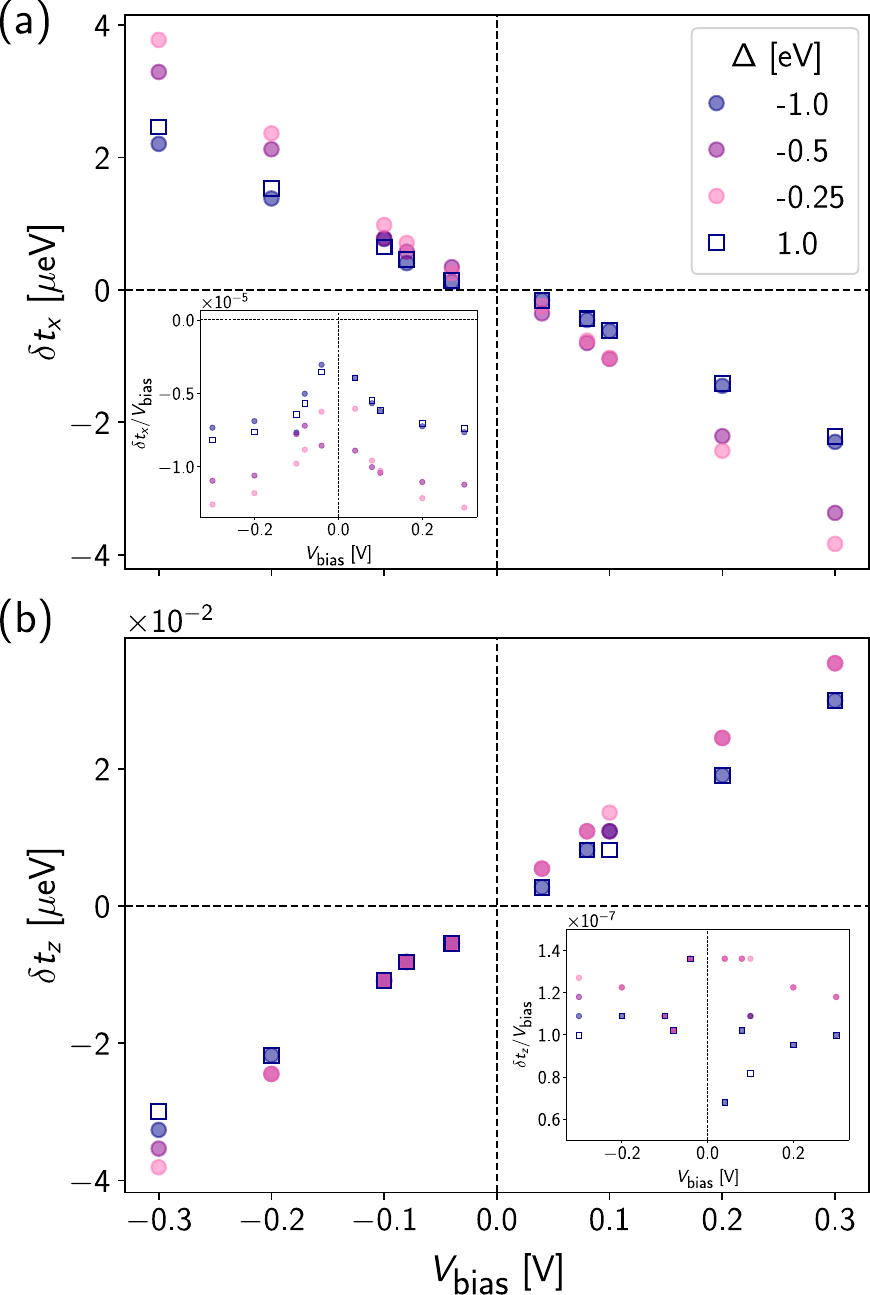}      
    \caption{Local SO torque response, $\delta t_\alpha$, exerted at the spin located in the vanadium atom as a function of the voltage bias, $V_\textnormal{bias}$, applied across the molecular junction. We display the traces obtained for several representative values of the magnetic anisotropy term in two relevant spatial directions (a) $\hat{\mathbf{x}}$ and  (b) $\hat{\mathbf{z}}$.
  Insets: ratio of the local SO torque reponse to the bias voltage in (a) $\hat{\mathbf{x}}$ and (b) $\hat{\mathbf{z}}$ direction.
    }\label{f5} 
\end{figure}

{\bf Results - torque.} 
Similar to the magnetization shown in Fig. \ref{f4}, we display in Fig. \ref{f5} the local SOT in (a) the $\hat{\mathbf{x}}$ and (b) the $\hat{\mathbf{z}}$ direction (see SI for the $\hat{\mathbf{y}}$ direction) for different values of $\Delta$. 
We first note that SOT is odd under time reversal and therefore 
changes sign upon reversing the bias voltage. 
This behavior is reflected in simple models employed for heterostructures utilizing the Rashba Hamiltonian in the presence of magnetization\cite{Gambardella2011}; in these instances it can be demonstrated explicitly that the SOT exhibits a direct proportionality to the current density. 
Also, the (linear) SOT is seen to be largely insensitive  to the magnetic anisotropy parametrized by $\Delta$, so the limit of interest, $\Delta\to 0$, is trivial to take. This extrapolation yields very small values for the ratio $\delta t_\alpha/V_\textnormal{bias}$ in the limit $V_\textnormal{bias} \rightarrow 0$. 
Further, owing to the definition \eqref{torqueopdev},  
$\hat{\mathbf{T}} = \bm{\sigma}\times\hat{\mathbf{B}}_{\textnormal{xc}}$, 
the magnitudes of torque and magnetization are proportional to each other. Correspondingly, 
the torque in the directions perpendicular to the magnetic moment exceeds the parallel torque (roughly in $\hat{\mathbf{z}}$-direction, Fig. \ref{f5}) by two orders of magnitude. 
%

{\bf Discussion - literature comparison.} 
In linear response, we have the relation 
$\mathbf{T}=\mathfrak{t}\mathbf{E}$ which introduces the torkance tensor $\mathfrak{t}$ and the probing electric field $\mathbf{E}$; the tensor is well accessible within the conventional linear-response (Kubo-) formalism and has been explored, e.g., in the context of inversion asymmetric magnetic heterostructures\cite{Freimuth2014, Geranton2015, Geranton2017}. 
For a  (semi-)quantitative comparison of our results obtained for the vanadium complex with such literature values, we offer a rough estimate: 

We first simplify the torkance tensor taking it to be diagonal with two non-vanishing eigenvalues, 
$t\coloneqq t_x=t_y$ corresponding to the two orthogonal directions $\hat{\mathbf{x}}$ and $\hat{\mathbf{y}}$. We further take the local electric field as orthogonal to the direction $\hat{\mathbf{z}}$ with magnitude
$E\approx V_\text{bias}/L$,  
where $L \sim 13 a_0$ is the length of the molecular layer measured as the linear distance between the two anchor groups in units of Bohr's radius $a_0$. 
Fig. \ref{f3} suggests that the linear response regime extends at least up to bias voltages of  $V_\textnormal{bias} \sim 40 $ meV, corresponding to an electric field $E \sim 3\, \textnormal{mV}/a_0$. 
Considering the ratio $t = |\mathbf{T}|/|\mathbf{E}|$, based on Fig. \ref{f5} we estimate $t\sim 3 \cdot 10^{-4} e a_0$. 
For a meaningful comparison with literature values obtained in cobalt (Co) heterostructures\cite{Freimuth2014}, we rescale the typical torkance values with the typical SO interaction energy, assuming a linear relation between the SOT and the SO interaction strength at lowest order. 
To estimate the SO interaction strength in the case of the molecular junction, we adopt the typical energy splitting of vanadocene molecular levels due to SO interaction, $\Delta E \sim 1$ meV, as obtained from the Kohn-Sham energies of our DFT calculations (\textcolor{black}{see Table 1 in the SI}); this results in $\xi_V = t/\Delta E \sim 0.3 a_0/\textnormal{V}$, for the vanadium-based molecule. 
For Co-based structures on the other hand, we obtain the rough estimate $\xi_\textnormal{Co} = t_\textnormal{Co}/\zeta_\textnormal{Co} \sim 1.25 a_0/\textnormal{V}$, where we combined a typical value for the torkance in the heterostructure active layer, $t_\textnormal{Co} \sim 0.1 e a_0 $, from Ref. \onlinecite{Freimuth2014} and a typical energy splitting associated with the SO coupling in a hydrogen-like atom, $\zeta_\textnormal{Co} \sim 0.123$ eV from Ref. \onlinecite{Bendix1993}. 
As one would expect, the SO-normalized results turn out to be of the same order of magnitude. 

{\bf Further discussion - parasitic magnetic fields.} 
The SOT can be (re-)interpreted as an effective (Oersted or coercive) magnetic field, 
$\mathbf{B}_\text{coerc} = \mathbf{T} \times \hat{\mathbf{m}}/\mu$, where $\hat{\mathbf{m}}$ denotes the initial direction of the magnetization before the SOT is applied and $\mu$ is the total magnetic moment of the interface molecule. 
\cite{Miron2010, Miron2011, Freimuth2014}. $\mathbf{B}_\text{coerc}$ represents the magnetic field strength that is needed to induce an equivalent effect on the magnetization as the current-induced SOT. 
Formally, the coercive magnetic field is a function of the applied bias, $\mathbf{B}_\text{coerc}(V_\text{bias})$, because the SOT depends on the current density.
Since in the vanadocene junction $\hat{\mathbf{m}} = \hat{\mathbf{z}}$, the coercive field is oriented perpendicular to the easy magnetization axis of 
$\hat{\mathbf{z}}$. For the $x-y$ plane, the typical value of SOT in the linear response regime is $\sim \mu$eV, which translates into $|\mathbf{B}_\text{coerc}| \sim 0.01$ Gauss. 
With an eye on the production of nano-structured magnetic fields in mesoscopic samples, we mention that  this value is two orders of magnitude smaller compared to typical current-induced magnetic fields in eddies in graphene sheets\cite{Walz2014}. Therefore, 
the SOT is likely to compete with unintended (``parasitic") current-induced magnetic fields that may impair a controlled SOT-mediated switching process, unless a system with large-enough SO coupling has been chosen. 


{\bf Conclusion.} 
We have presented a self-consistent calculation of SOT in a vanadocene-copper molecular junction at finite bias and in the steady-state regime.
Our findings are as follows: 
(a) the window of linear responses of current and torque is wide and extends beyond 300 meV. 
(b) Within this voltage window, the local magnetization being located near the vanadium atom is largely insensitive to the current flow. 
(c) The torque per voltage takes values that are roughly comparable with an earlier report of torkance coefficients for asymmetric Co-based  heterostructures \cite{Freimuth2014}. 
Our results suggest that magnetism might be controlled by current-induced SOT in single-molecule junctions. They also suggest, that generic spin-orbit coupling values are in a range so that SOT-mediated switching may compete with effects related to current-induced magnetic stray fields. Our results present a first step in the way to understanding microscopic SO-induced magnetism, torques and spin dynamics at the molecular level in tunnel junctions.

\subparagraph{Supporting Information.}
The Supporting Information is available free of charge at \texttt{http://pubs.acs.org}.\\
Computational details, technical mathematical derivations of the non-equilibrium density matrix with spin-orbit coupling, details on the DFT-NEGF cycle with spin-orbit coupling, relation between the expressions of the equilibrium and non-equilibrium density matrix, additional computational data of observables not shown in the main text, junction geometry.

\subparagraph{Acknowledgments.} 
The authors are thankful to E. K. U. Gross, R. Koryt\'ar, L. Kronik, J. J. Palacios, V. Pokorn\'y, and J. Wilhelm for helpful and inspiring conversations.
All the computations were carried out in the Athene Cluster at the University of Regensburg. 
This research was supported by the German Research Foundation (DFG) through the Collaborative Research Center SFB 1277 (Project-ID 314695032, subprojects A03, B01).
M.C.-G. is grateful to the Azrieli Foundation for the award of an Azrieli International Postdoctoral Fellowship.
D. H.-P. acknowledges funding from the Diputaci\'on Foral de Gipuzkoa through Grant 2023-FELL-000002-01.

\vspace{0.2cm}

\def\bibsection{}

\leftline{ \textbf{REFERENCES} }
\vspace{0.1cm}

\bibliography{biblio}

\end{document}